\documentclass[prd,showpacs,preprintnumbers,amsmath,amssymb,amsfonts,nofootinbib]{revtex4}
\everymath{\displaystyle}
\usepackage[dvips]{graphicx}
\usepackage{longtable}
\usepackage{subfigure}

\newcommand{\ide}{\mbox{1 \kern-.59em {\rm l}}}
\def\draftversion{Y}                
\def\note[#1]#2{\message{(#1)}\if\draftversion
{\noindent\em[#2]\/}\fi}
\if \draftversion N
\fi
\preprint{IMSc/2007/06/7}
\begin{document}
\title{Finite temperature phase transition of a single scalar field
on a fuzzy sphere}

\author{C. R. Das}
\email{crdas@imsc.res.in}
\author{S. Digal}
\email{digal@imsc.res.in}
\author{T. R. Govindarajan}
\email{trg@imsc.res.in}
\affiliation{The Institute of Mathematical Sciences, C.I.T. Campus,
Taramani, Chennai  600 113, India}

\begin{abstract}
We study finite temperature phase transition of neutral
scalar field on a fuzzy sphere using Monte Carlo simulations. 
We work with the zero mode in the temporal directions, while
the effects of the higher modes are taken care by the temperature
dependence of $r$. In the numerical calculations we use 
``pseudo-heatbath'' method which reduces the auto-correlation 
considerably. Our results agree with the conventional calculations. 
We report some new results which show the presence of meta-stable 
states and also suggest that for suitable choice of parameters 
the symmetry breaking transition is of first order.

\end{abstract}

\pacs{12.60.Rc; 12.10.-g; 14.80.Hv; 11.25.Wx; 11.10.Hi}

\keywords{ Non-commutative geometry}

\maketitle

\section{Introduction}

QFT's on non-commutative spaces have been studied from various perspectives
recently \cite{doug-nek,szabo,nair,connes,szabo2,sachin,trg1,trg2}.
Most frequently studied NC space is the well known Groenwald Moyal space 
\cite{moyal}
$R_\theta^{2d}$ and various issues like, renormalisation, causality, solitons,
statistics have been analysed in the literature
\cite{minwalla,gopakumar,chaichian,greenberg}.
The conventional quantisation of fields
on these spaces have led to an interesting behavior known as IR/UV mixing.
The phase structure of fields on such a space reveals a new phase known as
strip phase \cite{gubser}.
Alternative quantisation which preserves a twisted Poincare symmetry in these
theories avoids such a difficulty \cite{bal,baltrg}.

On the other hand the fields on fuzzy spaces like fuzzy spheres, fuzzy $CP_n$ etc
are explicitly finite and do not have the IR/UV mixing \cite{vaidya,madore}.
But there is an anomaly in the finite case which reveals itself as generating the
IR/UV mixing. There is lot of confusion about taking the limit of continuum in these
models and it has been pointed out various possibilities do exist 
\cite{denjoe,steinaker,madore,vaidya}.

The QFT on fuzzy sphere is a matrix model and easily amenable to simulations and
numerical studies \cite{xavier,bitenholz,panero}.
We study a real scalar field on the fuzzy sphere using Monte Carlo
simulations. The earlier studies  involved metropolis
algorithm to ensure the randomness of fluctuations but the autocorrelations are
reduced using over-relaxations \cite{panero}. But we will use another
technique extensively used in the study of Higgs model - known as
pseudo-heatbath \cite{bunk}. Using this algorithm we have been able to 
reproduce previous results from different studies. Apart from this we
are able to characterise the order of the transitions between the different 
phases. In particular we find the transition between order $\leftrightarrow$
non-uniform transition is of first order which is mostly due to the presence of many 
meta-stable states at low temperatures in the model. Also we find new results for 
the structure of the phase diagram as well as for scaling of the location of the triple 
points in the continuum limit.

The paper is organised as follows: the sec(2) introduces
the QFT on fuzzy spheres; sec(3) discusses the pseudo-heatbath technique
and brings out its salient features. Sec(4) reproduces the known results
and discusses our results on the nature of phases and the transitions.
In sec(5) we present our conclusions.

\section{The Model}

We use the following action for the massive, neutral, scalar field on the
fuzzy sphere of radius $R$ \cite{panero,xavier,denjoe},
\begin{equation}
S(\Phi) = \frac{4\pi}{N}{\rm Tr}\left[
\Phi\left[L_i,\left[L_i,\Phi\right]\right]+R^2\left(r\Phi^2+\lambda\Phi^4\right)
\right].
\end{equation}
\noindent Here $\Phi\in {\rm Mat}_N$ is a $N\times N$ hermitian
matrix. The first term in the action is (kinetic) coming from
the variation of the $\Phi$ field on the fuzzy sphere. The quartic
term represents the self interaction of the $\Phi$ field. For the thermal
behavior of the $\Phi$ field one needs to study the system in $2+1 d$. However
one can consider
the above action as the dimensionally reduced version of a $2+1$ dimensional
action with the effects of temperature going into the temperature
dependence of $r$. Finite temperature behavior of the $\Phi$ field
with fluctuations included is then studied at different $r$. In the mean field 
approximation the expectation value of $\Phi$ field which minimises the action
is given by, 

\begin{equation}
\langle\Phi\rangle = \pm \phi \ide,\quad \phi= \sqrt{-{r \over 2\lambda}}
\end{equation}

\noindent $\phi \ge 0$ for negative values of $r$ and zero for $r \ge 0$.
The system is in ordered phase for $r < 0$ and in the disorder phase
for $r \ge 0$. So at the mean-field level there are only two phases.
$\langle\Phi\rangle$ decreases continuously to zero in the limit
$r \rightarrow 0$. At $r=0$ the system undergoes a second order phase 
transition with mean field critical exponents, $\beta=1/2, \alpha=0$ etc.. In 
the disorder phase $\langle\Phi\rangle =0$ so the $Z_2$ symmetry of the model 
is restored. Even though the above form of $\langle\Phi\rangle$ minimises the 
action there are additional local minima or meta-stable states. For these 
states, form of $\langle\Phi\rangle$ is non-identity in general. Some of these
states will have same potential as the ground state. In the mean-field 
approach these states do not play any role in the phase transition. 
However they become important when fluctuations
beyond mean-field are considered.

Hence the next step in the calculations is to consider effect of thermal 
fluctuations beyond mean field. It is important then to ask if the results 
of the mean field analysis survive. One expects that the fluctuations will 
destroy a non-zero $\langle\Phi\rangle$ even for $r$ less than the mean-field
critical value which is zero. Further more the fluctuations may lead to 
new phases and different types of phase transitions.
These are some of the issues of intense numerical investigations recently 
\cite{panero,xavier,denjoe}. So far the results show the appearance of a new 
phase called non-uniform ordered phase. These studies are mostly done using a 
standard Monte Carlo simulations with metropolis algorithm. 

In the present work we study the fluctuations in the above model using a 
different numerical technique known as ``pseudo-heatbath'' method. Like the 
previous studies we also observe the non-uniform phase. However our results 
seem to indicate that there are phase structure within the non-uniform phase. 
These phases can be probed using different operators/order parameters. As a 
consequence there will be multiple triple points in the $\lambda -r $ plane. 
In the following we describe the ``pseudo-heatbath'' method. Subsequently we 
present and discuss our results next section.

\section{Numerical Technique}

Effects of the fluctuations beyond mean field are computed from
the partition function, which in the path integral approach
is given by,

\begin{equation}
{\cal Z}\; \propto\; \int D\Phi e^{-S(\Phi)}.
\end{equation}

The standard numerical methods adopted for this
integration are Monte Carlo simulations. In the Monte Carlo algorithms,
one generates an ``almost'' random sequence of $\Phi$ matrices by
successively updating elements of $\Phi$ taking into account the measure 
and the exponential in the integral above. This sequence
of $\Phi$ is then used as an ensemble for calculating averages of various
observables. For example, thermal average of $\Phi$ is given by,
\begin{equation}
\langle \Phi\rangle = {1 \over N_m}\sum_{m=1}^{N_m}\Phi_m,
\end{equation}
here, $\Phi_m$ is the $m$th element of the ensemble. Usually there are
different ways to generate the ensemble. Previous studies of this model have 
considered the metropolis algorithm \cite{xavier,panero}.
In the metropolis updating usually there is a substantial correlation
between $\Phi$'s in the sequence. For a good ensemble the auto correlation
between the configurations in the sequence must be really small. Though this 
auto correlation can be reduced by using some over relaxation programme 
\cite{panero}. The auto-correlation is greatly reduced, however, when 
``heatbath/pseudo-heatbath'' type of algorithms are used \cite{bunk}. This 
method is very much common in the non-perturbative study of $\Phi^4$ theories 
in conventional lattice simulations. It gives better sampling and is efficient 
at least for smaller $\lambda$ values. This is why we make use of 
``pseudo-heatbath'' technique. In the following we explain the algorithm
in greater detail.

In the ``pseudo-heatbath'' algorithm, given a $\Phi$ we update 
the elements of this matrix one at a time. Advantages of updating matrix
elements were demonstrated in Ref. \cite{bholz}. 
Keeping in mind that $\Phi$ is hermitian we update $\Phi_{ij}$ and $\Phi_{ji}$ 
simultaneously. We update $\Phi_{ij}$ using the probability distribution,
\begin{eqnarray}
P\left(\Phi_{ij}\right) &=& e^{-S\left(\Phi_{ij}\right)}\nonumber\\
{\rm where}\qquad S\left(\Phi_{ij}\right) &=& \alpha\left(\Phi_{ij}-A\right)^2
+\lambda B\left(\Phi_{ij}-C\right)^4,
\end{eqnarray}
\noindent $A$, $B$, $C$ may depend on the elements of $\Phi$ (other
than $\Phi_{ij}$). $\alpha$ is a parameter chosen so as to maximise
the efficiency of updating. In the
first step a random number is generated
using the distribution,
\begin{equation}
e^{-\alpha\left(\Phi_{ij}-A\right)^2}.
\end{equation}
\noindent In the second step the newly generated random number is
accepted or rejected using the second term of $S\left(\Phi_{ij}\right)$.
In our calculations we get for some choice of the
parameters, in particular small $\lambda$ acceptance rate up to $95\%$.
Over relaxation can also be easily incorporated into this algorithm. In 
the over relaxation process we flip the element $\phi=\Phi_{ij}$ in the
following way,
\begin{equation}
\phi^\prime = A - 2\phi
\end{equation}
\noindent then accept it with the probability $\exp(-\delta S)$. $\delta S$
is the change in action due to flipping. For small
$\lambda$ this amounts to changing $\Phi$ by large amount with only a
small change in the total action. Even without using the over relaxation
method we get very small auto correlation. In the following we present and 
discuss our numerical results.

\section{Numerical Results and discussion}

To study the phase diagram and transitions we make measurements of various
observables such as
\begin{eqnarray}
Tr(\Phi),\quad Tr(\Phi^2),\quad Tr(S)
\end{eqnarray}
\noindent at various values $r R^2$ for different choices of $(N, \lambda
R^2)$. In order to check our algorithm we considered some of the 
parameters used in previous calculations \cite{panero,xavier,denjoe}, and 
found that our results match reasonably well with previous results. 
The results also agreed with mean-field away from the transition point. In 
Fig.~1(a) we show Monte Carlo history of $Tr(\Phi)$ for $N=2$, $R=1.0$,
$\lambda=0.63662$ and $r=-1.530502$ as used in \cite{panero}. Without using 
the over relaxation we get the quality of data similar to that of 
\cite{panero} shown in Fig.~1(b). The average values agree but we
observe larger fluctuations (Fig.~1a).

\begin{figure}[hbt!]
\begin{center}
\subfigure[\; Monte Carlo history of $Tr(\Phi)$.]
{\label{fig1a}\includegraphics[scale=0.65]{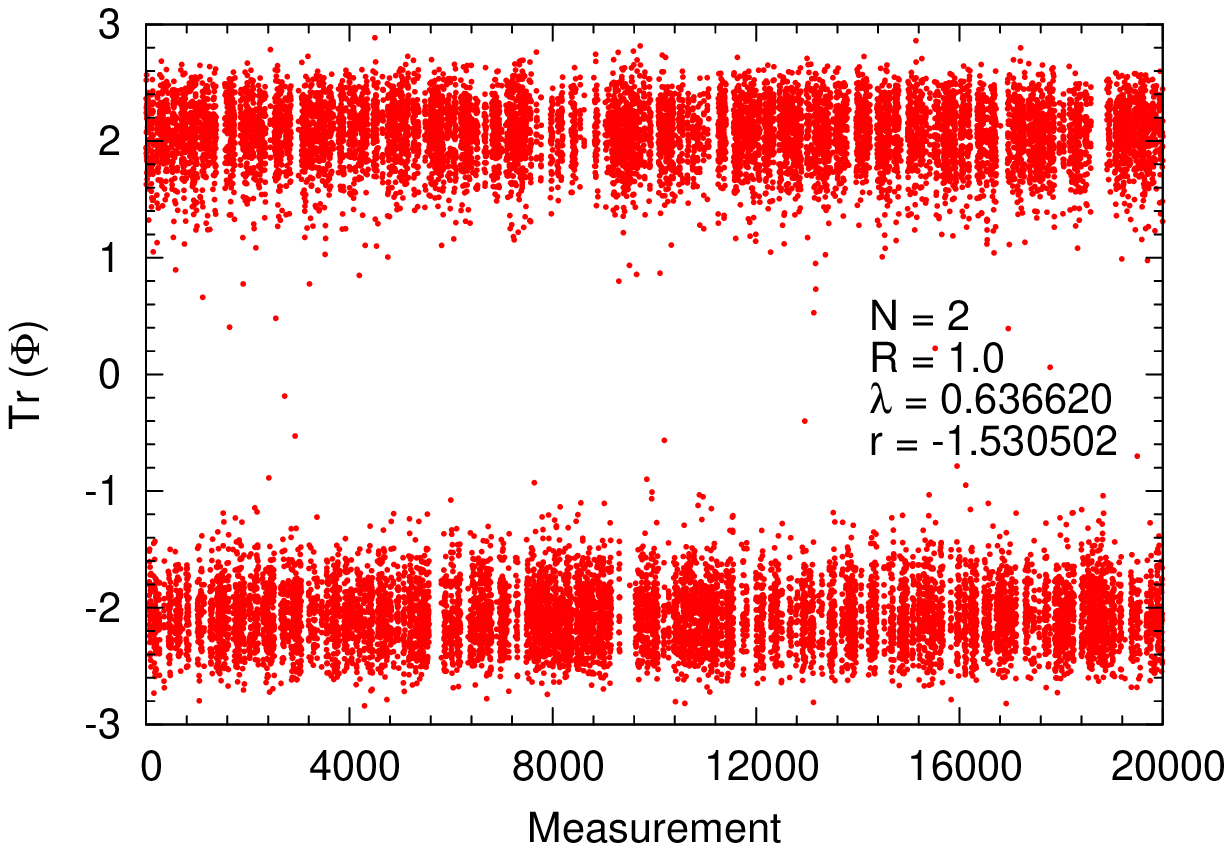}}
\subfigure[\; Monte Carlo history of $Tr(\Phi)$ from \cite{panero}.]
{\label{fig1b}\includegraphics[scale=0.375]{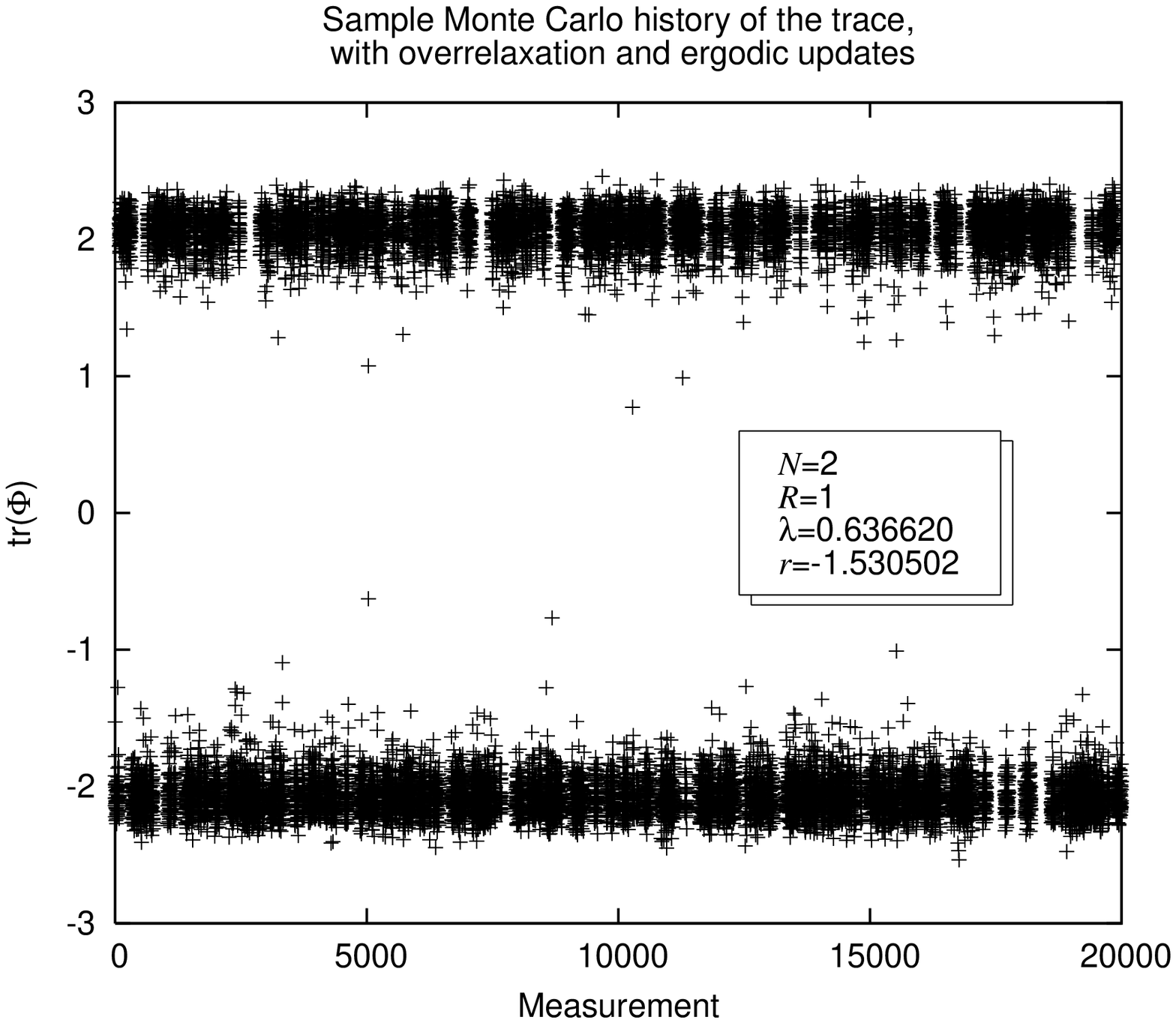}}
\caption{Comparison between pseudo-heatbath method and Metropolis method}
\label{fig1}
\end{center}
\end{figure}

\vskip0.25cm

\noindent{\bf Order $\leftrightarrow$ non-uniform transition}\\

Having reproduced some of the previous results we considered
various values of the parameters $(N, \lambda R^2)$ to study the
phase diagram. For large
values of $\lambda R^2$ we found multiple transitions \cite{denjoe}. For 
low temperatures, or $r \ll 0$, the average value of $\Phi$ is essentially 
identity matrix indicating the ordered phase. For larger values
of $r$ the average of $Tr(\Phi)$ vanishes while the average of some elements
$\Phi_{ij}$ is non-zero. Such a form of $\Phi$
average indicates a non-uniform phase which breaks  spatial rotation
spontaneously. Since the average of $Tr(\Phi)$ is non-zero in the
order phase and zero in the non-uniform phase one can use it as an
order parameter for the order $\leftrightarrow$ non-uniform transition.

As for the order of the non-uniform $\leftrightarrow$ order transition 
we find it is strong first order for larger $\lambda R^2$. This can be 
seen from the hysteresis effects of $Tr(\Phi)$. In Fig.~2(a) we show the 
hysteresis loop of $Tr(\Phi)$ for the set of parameters $N=25$ and 
$\lambda=0.8$. The value of $r$ corresponding to the middle of
the hysteresis loop is take to be the critical(transition) value
$r=r_1$ for this transition. By doing simulations for different values 
of $\lambda$ we find that the strength of this 
order $\leftrightarrow$ non-uniform transition varies. For smaller
$\lambda$ the transition becomes weaker. For example, for $\lambda=0.4$
the hysteresis loop was not prominent.

\begin{figure}[hbt!]
\begin{center}
\subfigure[\; Hysteresis loop of $Tr(\Phi)$ for
$N=25$, $R=10.0$ and $\lambda=0.8$.]
{\label{fig2a}\includegraphics[scale=0.595]{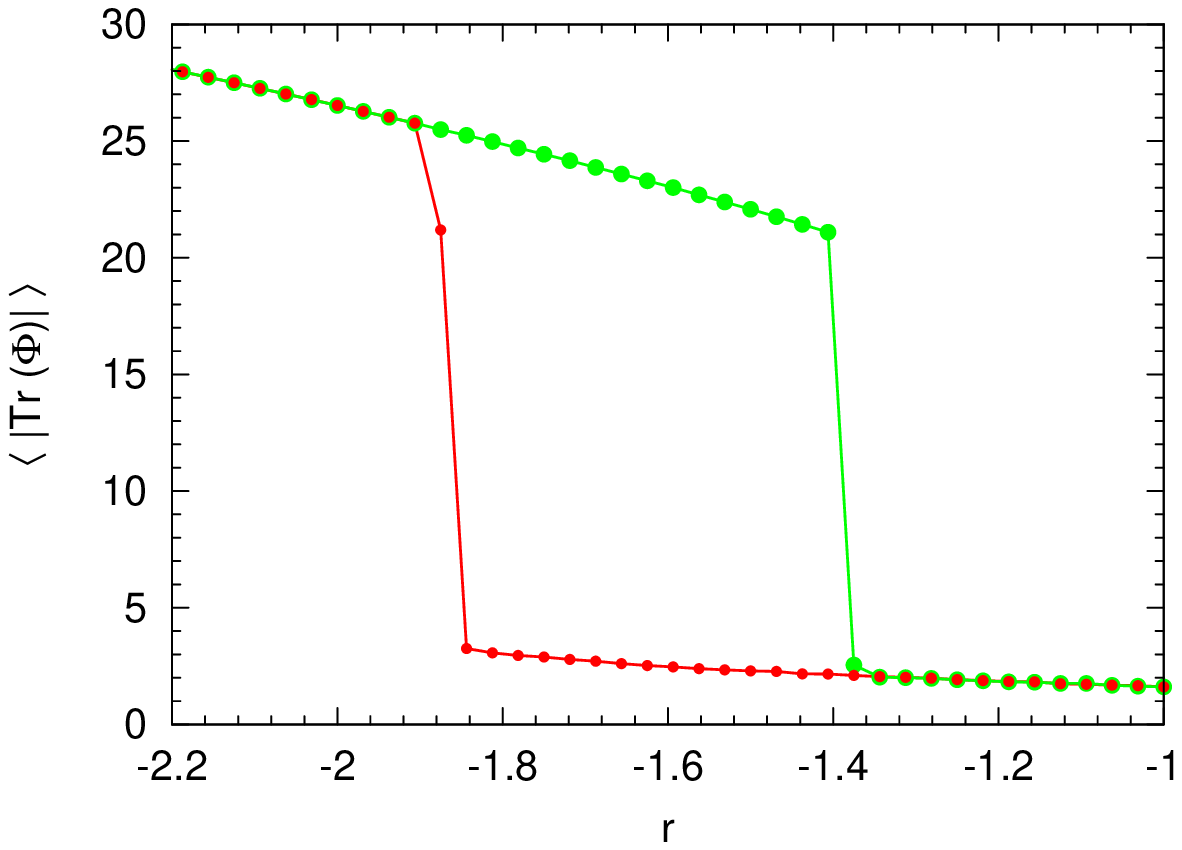}}
\subfigure[\; Monte Carlo history of $Tr(\Phi)$ for 
$N=25$, $R=10.0$, $\lambda=0.4$ and $r=-0.75$.]
{\label{fig2b}\includegraphics[scale=0.595]{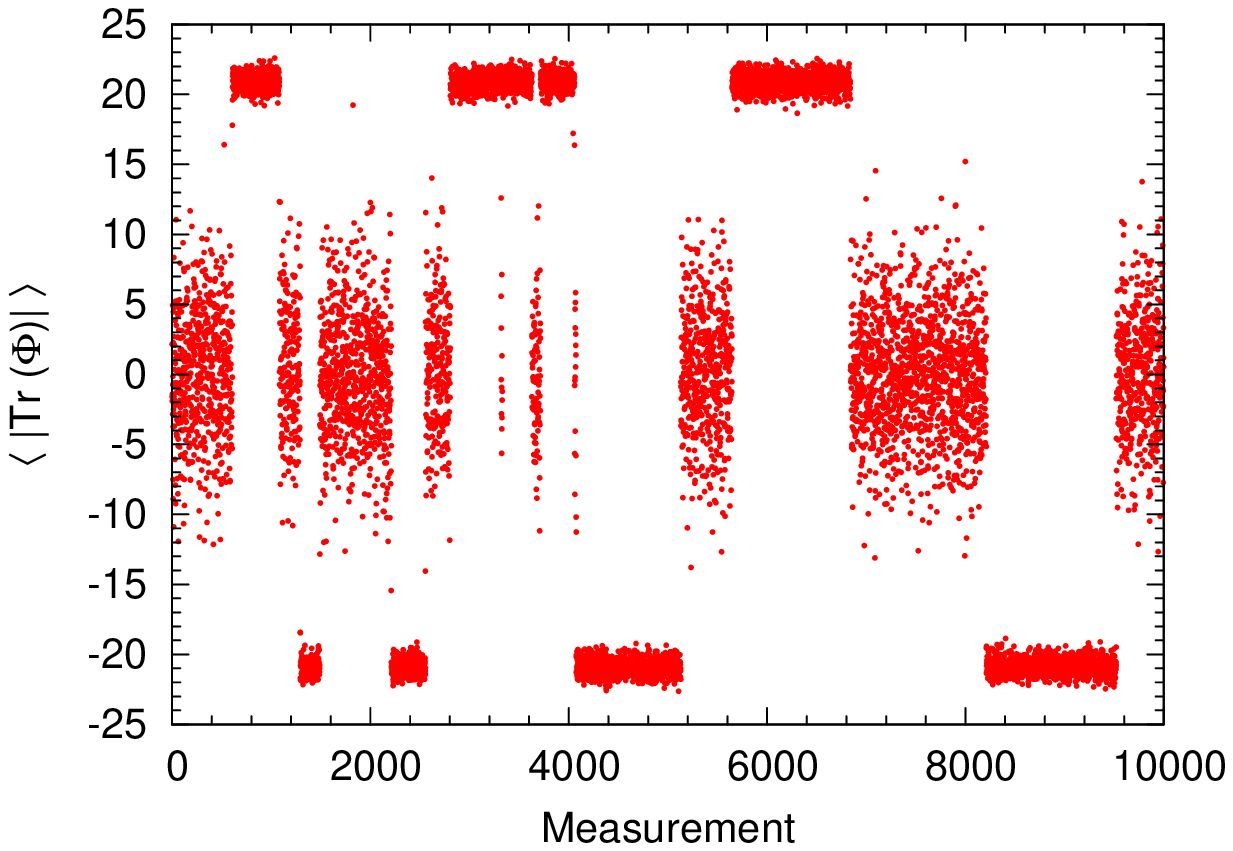}}
\caption{First-order Phase Transition}
\label{fig2}
\end{center}
\end{figure}

The history of measurement of $Tr(\Phi)$ and its distribution is shown in 
Fig.~2b for $r$ close to the corresponding 
critical value $r_1$. One clearly sees three 
degenerate ground states here. Out of these two are connected by $Z_2$ 
symmetry and the 3rd has $Tr(\Phi)$ peaked around zero and 
represents a non-uniform ordered phase. This distribution indicates that the 
order and non-uniform phases do coexist suggesting first order nature of
the transition between these two phases. Even though the transition is first 
order, it is weaker compared to the previous example. 
\vskip0.2cm
\noindent{\bf Non-uniform $\leftrightarrow$ disorder transition}\\

In the non-uniform phase $Tr(\Phi)$ keeps
fluctuating around zero. One can clearly see measured values of
$Tr(\Phi)$ form bands symmetrically situated around zero as shown in
figure Fig.~3(a). The band structure are not seen in the Monte Carlo history
of $Tr(\Phi^2)$ and $S(\Phi)$. This band structure of $Tr(\Phi)$ we saw 
mostly in the case when there were many meta-stable states before 
transition in the ordered phase, i.e for larger $N$. In Fig.~3(b) we
show the histogram of $Tr(\Phi)$ which clearly shows a peak close to
zero. This implies the state with lowest $Tr(\Phi)$ is the ground state of
the system and other bands are meta-stable states. 
The meta-stable bands tend towards zero as we increase $r$ further.
At the same time some bands disappear and/or others merge with the middle one.

\begin{figure}[hbt!]
\begin{center}
\subfigure[\; Monte Carlo history of $Tr(\Phi)$ at $N=16$, $R=15.0$,
$\lambda=0.7$ and $r=-0.8$.]
{\label{fig3a}\includegraphics[scale=0.595]{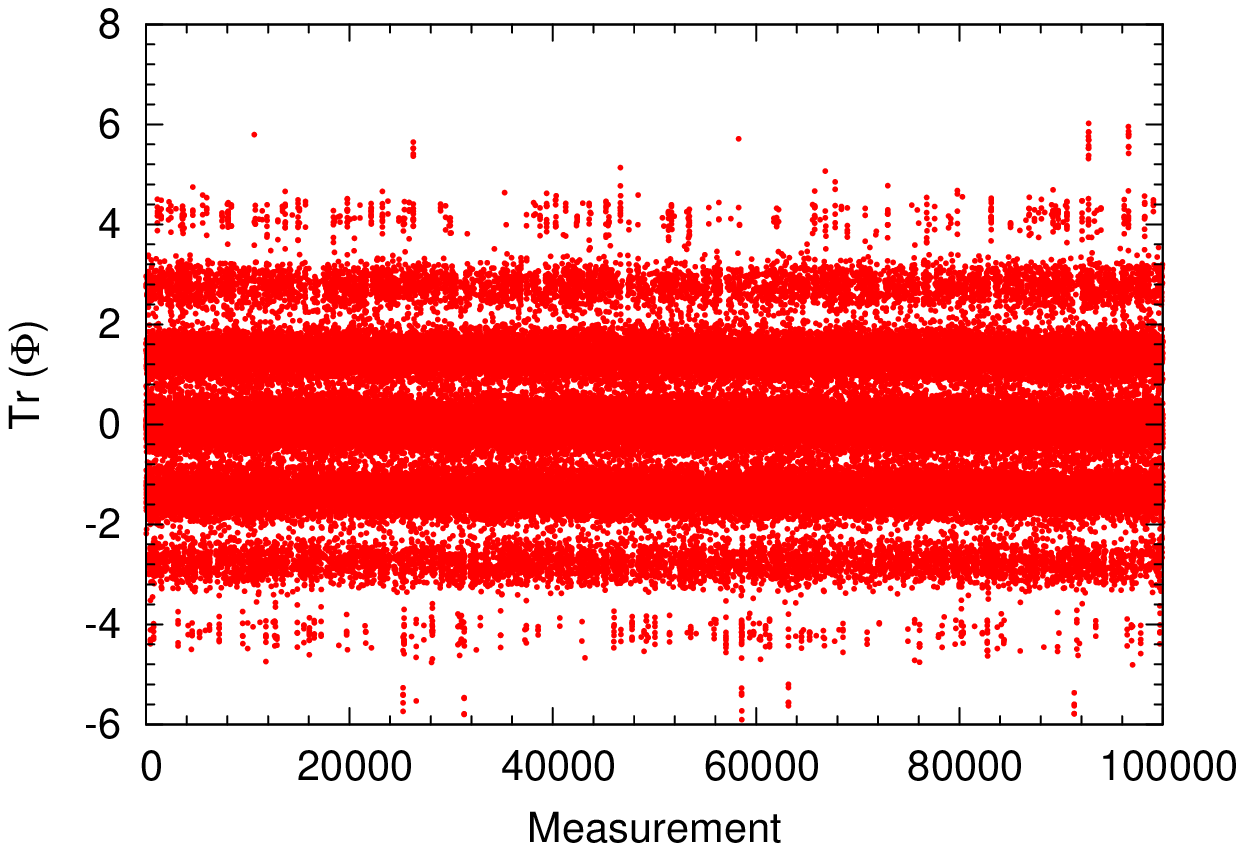}}
\subfigure[\; Histogram of $Tr(\Phi)$ at $Tr(\Phi)$ at $N=16$, $R=15.0$,
$\lambda=0.7$ and $r=-0.8$.]
{\label{fig3b}\includegraphics[scale=0.595]{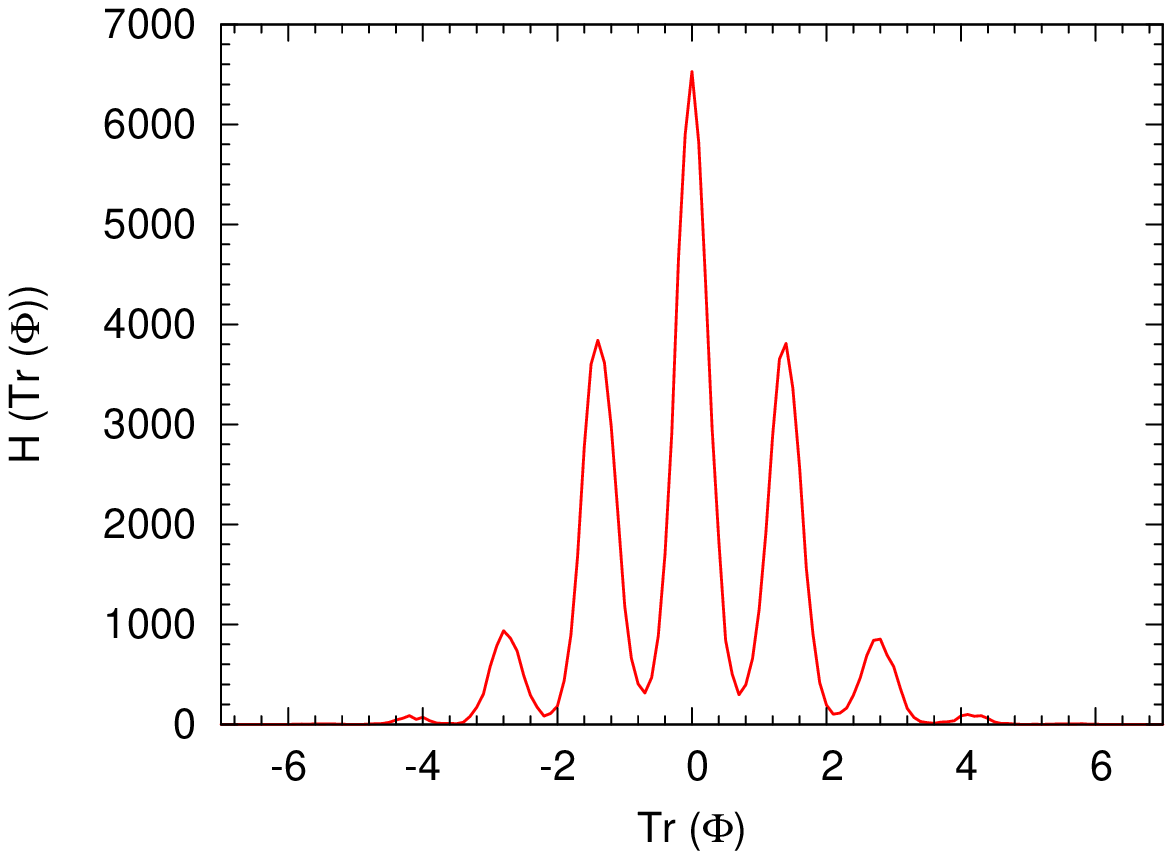}}
\caption{Monte Carlo history and histogram for $r_1 < r < r_2$ }
\label{fig3}
\end{center}
\end{figure}

In the basis we choose to work with even though $Tr(\Phi)$ fluctuates
around zero both $\Phi_{11}$ and $\Phi_{NN}$ fluctuate around non-zero
$Z_2$ symmetric values. So the symmetry of $\Phi$ is not restored 
yet. When we increase $r$ the
non-zero values around which the first and last diagonal of $\Phi$ 
fluctuate approach smoothly to zero. Beyond certain value of $r=r_2$ all 
elements of $\Phi$ fluctuate around zero restoring the $Z_2$ symmetry. 
In Fig.~4 we show the distribution of these elements of $\Phi$ both below 
and above $r_2$. Given this behavior of $\Phi_{11}$ and $\Phi_{NN}$ one can 
consider any of them as the order parameter for the 
non-uniform $\leftrightarrow$ disorder transition. This also implies that the 
higher spherical harmonics are becoming important for this transition in 
our basis \cite{denjoe}. When $r_2$ is approached
from below the peaks of the distribution of $\Phi_{11},\Phi_{NN}$ smoothly 
approach zero indicating only one ground state at any particular value of 
$r$ around the transition point. So this transition is a continuous
transition. 

\begin{figure}[hbt!]
\begin{center}
\subfigure[\; Distribution of $\Phi_{11}$ for $N=12$, $R=10.0$ and 
$\lambda=0.04$.]
{\label{fig4a}\includegraphics[scale=0.595]{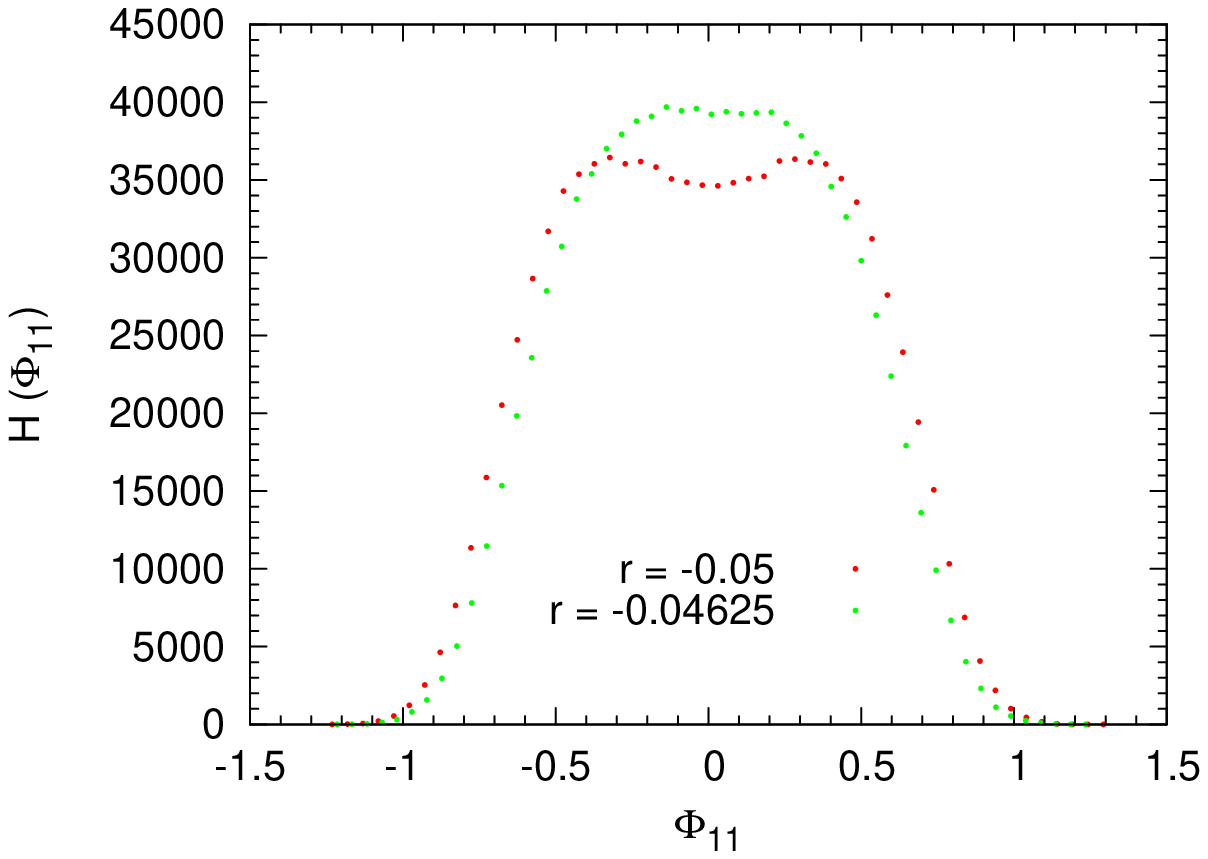}}
\subfigure[\; Distribution of $\Phi_{NN}$ for $N=12$, $R=10.0$ and 
$\lambda=0.04$.]
{\label{fig4b}\includegraphics[scale=0.595]{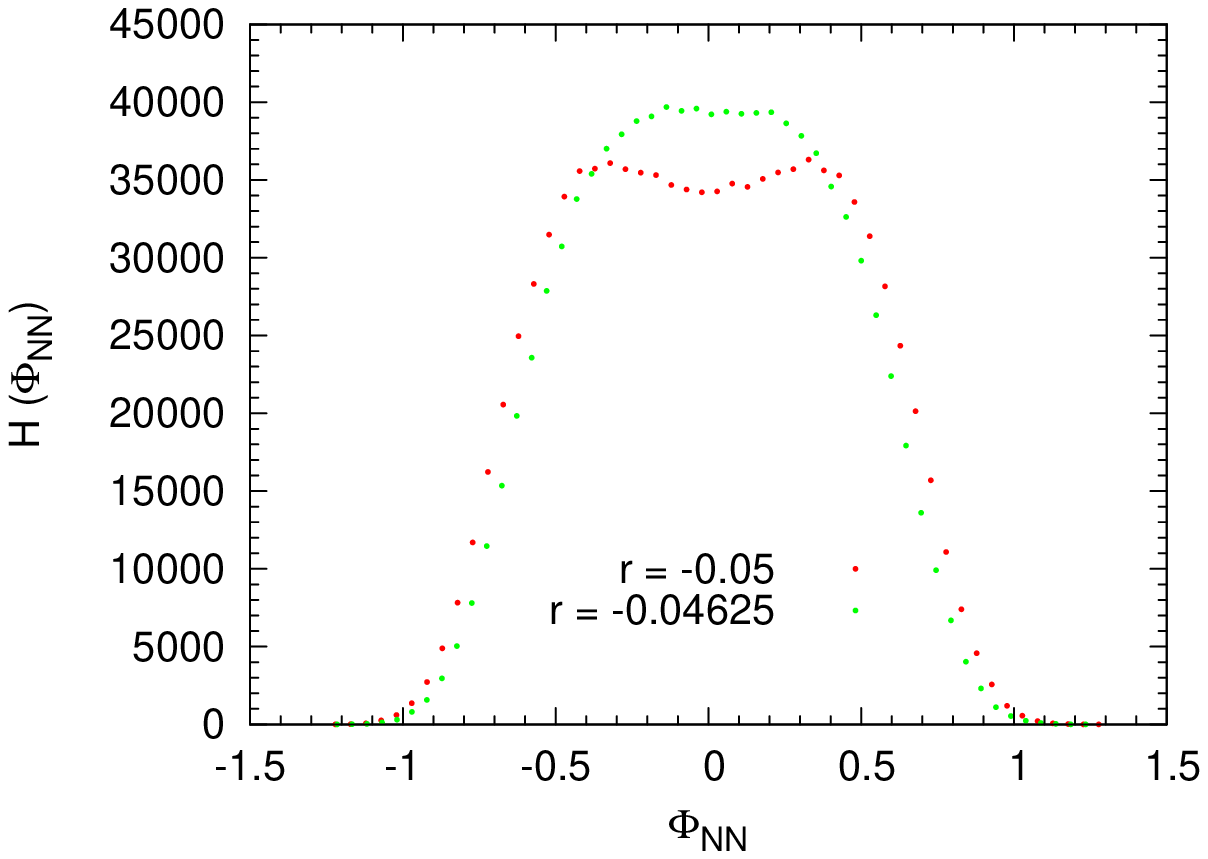}}
\caption{Non-uniform $\leftrightarrow$ disorder transition}
\label{fig3}
\end{center}
\end{figure}
\vskip0.25cm
\noindent{\bf Phase diagram and triple points}\\

When we analysed the data for fluctuations of $Tr(\Phi^2)$ we found these 
peaked at a certain value of $r$ within range $r_1 < r < r_2$ (Fig.~5b). 
This suggests finer structure or phases in the non-uniform phase. The peak
in $Tr(\Phi^2)$ then corresponds to the transition between these phases.

\begin{figure}[hbt!]
\begin{center}
\subfigure[\; Susceptibility of $Tr(\Phi)$ for $N=25$, $R=10.0$ and $\lambda=0.9$.]
{\label{fig5a}\includegraphics[scale=0.595]{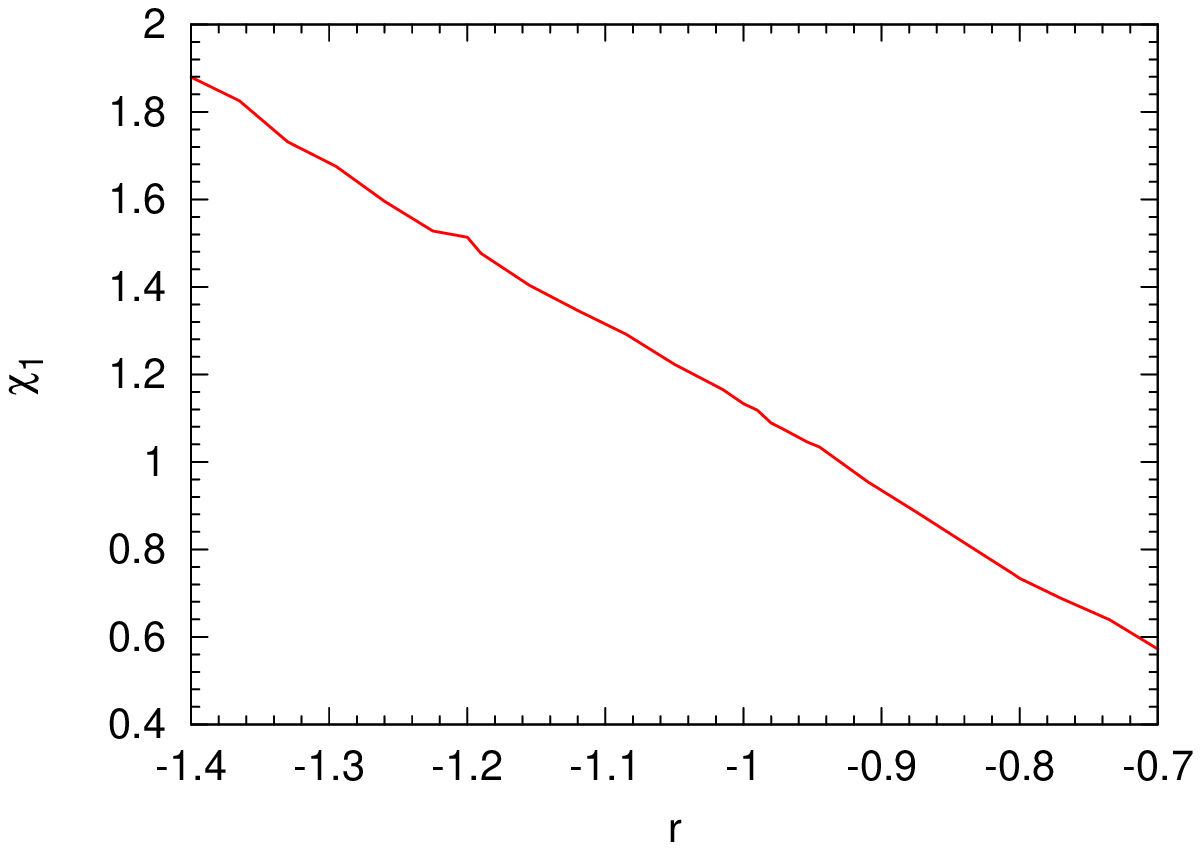}}
\subfigure[\; Susceptibility of $Tr(\Phi^2)$ for $N=25$, $R=10.0$ and $\lambda=0.9$.]
{\label{fig5b}\includegraphics[scale=0.595]{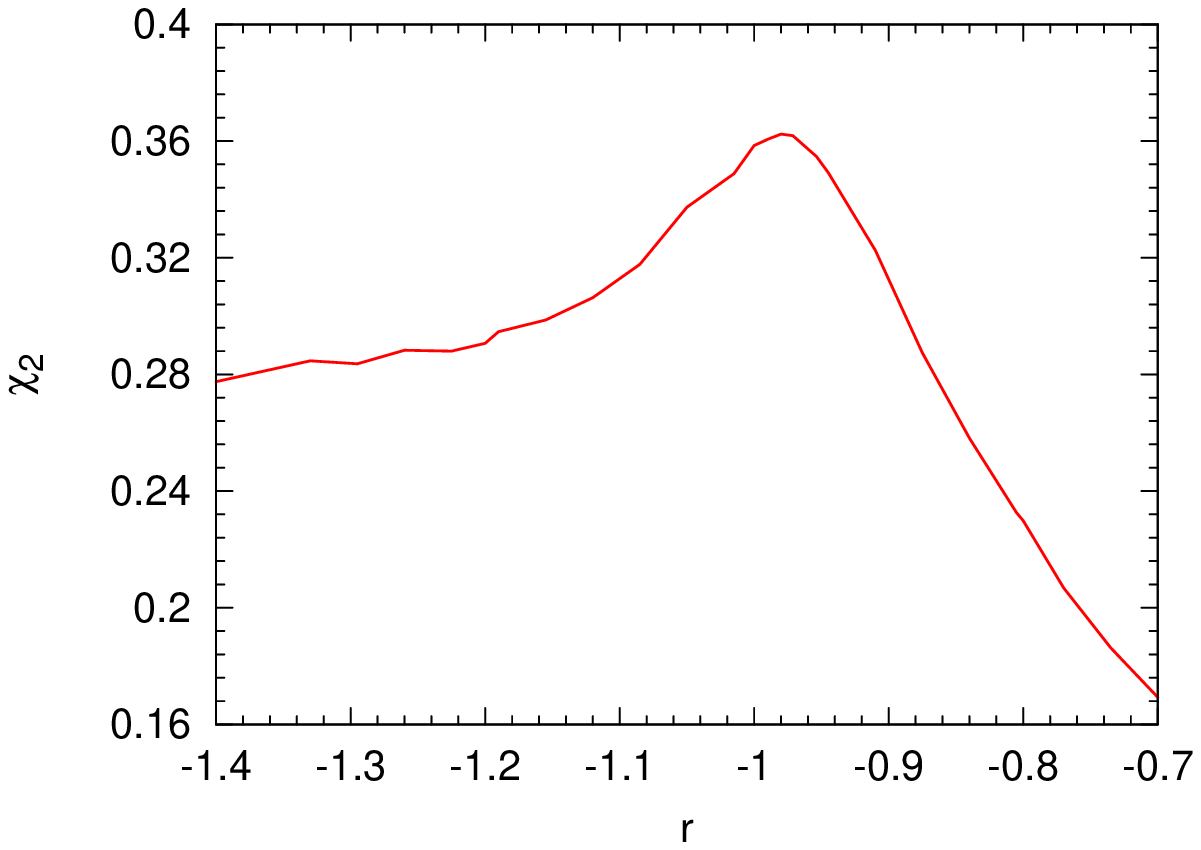}}
\caption{Transition between non-uniform phases}
\end{center}
\end{figure}

We did not see any dramatic change in the variables such as $Tr(\Phi)$ or
the diagonal elements of $\Phi$. The finiteness of the peak implies that
the transition is a continuous transition. For larger 
$\lambda$ the transition point was far from both $r_1$ and $r_2$. 
We anticipate that the non-uniform phase has more structure than what we see
from the behavior of $Tr(\Phi^2)$. This finer structure could be explored
by appropriate operators such as multi-trace operators. Note that this fine
structure becomes more prominent for larger $N$, which implies that
it will survive in the continuum non-commutative limit.

As $r$ is
increased from some large negative value the system explores all these phases
for large $\lambda$. For smaller $\lambda$ some
of these phases will not appear when $r$ is varied. This leads to presence of 
triple points in the $\lambda - r$ plane. For some small $\lambda$ there is 
transition directly between order $\leftrightarrow$ disorder phases, i.e 
$|r_1-r_2|$ vanishes. This leads the triple point 
which has the lowest value of $\lambda R^2$.
In Fig.~6a we show the phase diagram for $N=25$ in the $\lambda R^2~vs~r R^2$ 
plane. This is the triple point studied in previous works. In these studies the 
triple point was obtained by using numerical results for the order-non-uniform 
transition and the analytic results which takes into account only the potential 
term \cite{denjoe}. In our case both transitions lines are from our simulations.

Conventional lattice regularization of the
model does not show any evidence of non-uniform ordered phase. So it is
imperative to study what happens to the non-uniform ordered phase
in the continuum limit. If the non-uniform phase survives this limit 
then only it can be physically relevant. We have studied the $N$ dependence
of the triple point. Since it's not practical to do simulations for very 
large $N$, one must study scaling to find out 
the limiting position of the triple point for larger $N$ values. 
In the Fig.~6b we show value of $Y_{tri}=\lambda R^2, X_{tri}=r R^2$ 
corresponding to the triple point for different $N$.
The values of $N$ considered for our simulations are $N=4,~8,~12,~16, ~25$. 
When $N$ was increased the triple point moved away from the origin. Our
results suggest that the triple point scales with $(N^\mu,N^\nu)$ with 
$\mu\simeq 1.0 \simeq \nu$. We did not observe any universal scaling 
of the phase boundaries in the phase diagram for different $N$.

\begin{figure}[hbt!]
\begin{center}
\subfigure[\; Phase diagram for $N=25$.]
{\label{fig6a}\includegraphics[scale=0.595]{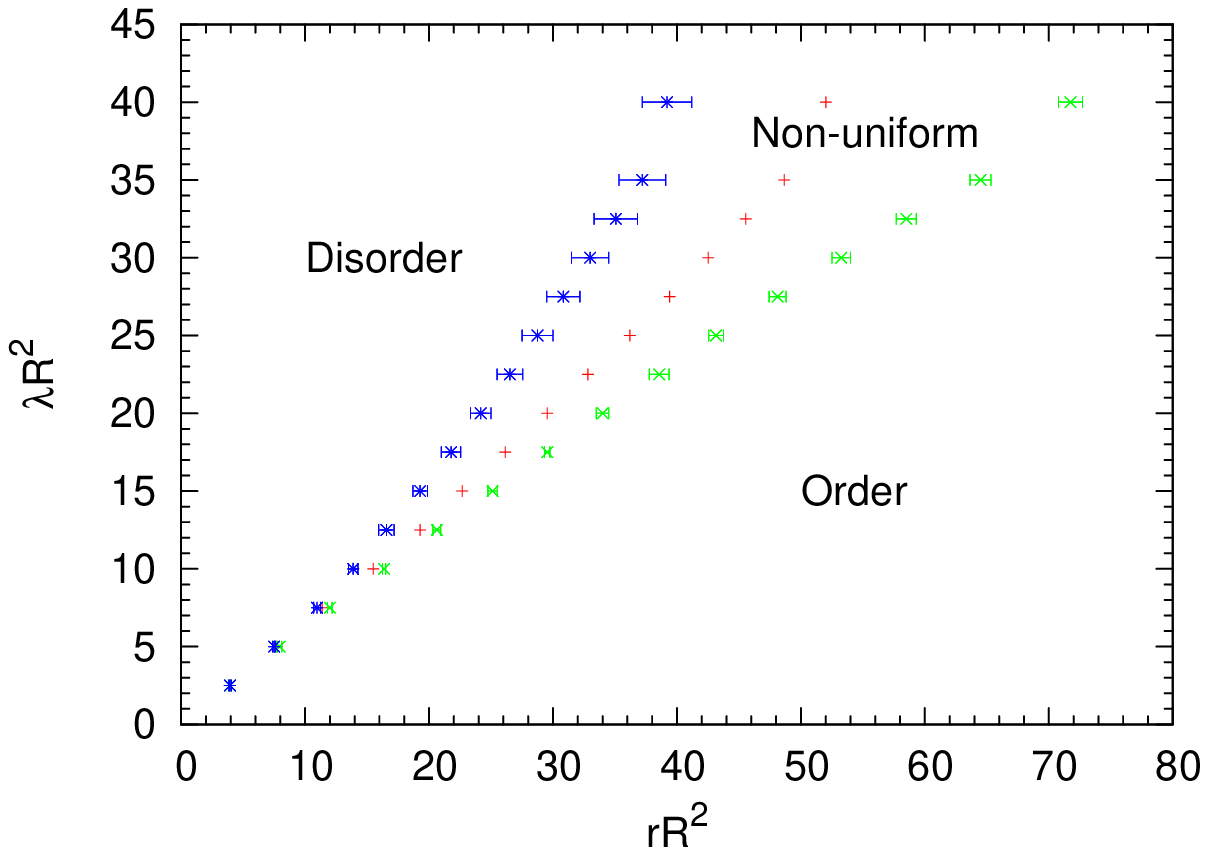}}
\subfigure[\; Scaling of the triple point with $N$.]
{\label{fig6b}\includegraphics[scale=0.595]{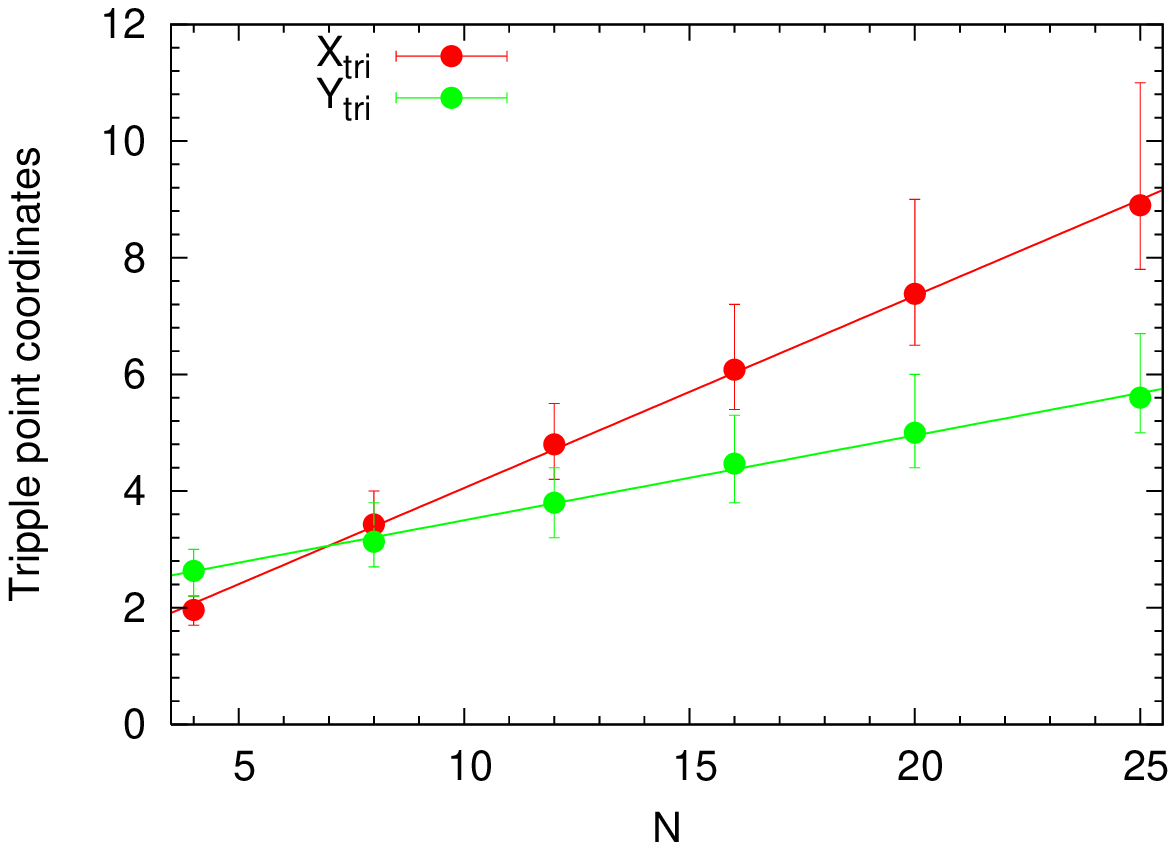}}
\caption{Phase diagram and scaling of the triple point}
\end{center}
\end{figure}

For smaller $\lambda R^2$ there is only one transition, 
the order $\leftrightarrow$ disorder transition. For larger $N$ the 
distribution of the observables such as $Tr(\Phi), \Phi_{11}, \Phi_{NN}$, 
close to the critical point, show a plateau around zero with highly 
non-gaussian features. 
This can be seen in Fig.~7 where have shown the histogram of $Tr(\Phi), 
\Phi_{11}$ around the critical
temperature. The parameters considered here are 
$\lambda=0.4,~R^2=10^2,~N=12$. We take 
the plateau structure around zero as an indication of a transition
which is stronger than second order transition. However it does not rule out
the possibility of second order phase transition for smaller values of 
$\lambda$ and $N$.

\begin{figure}[hbt!]
\begin{center}
\subfigure[\; Distribution of $Tr(\Phi)$]
{\label{fig3a}\includegraphics[scale=0.595]{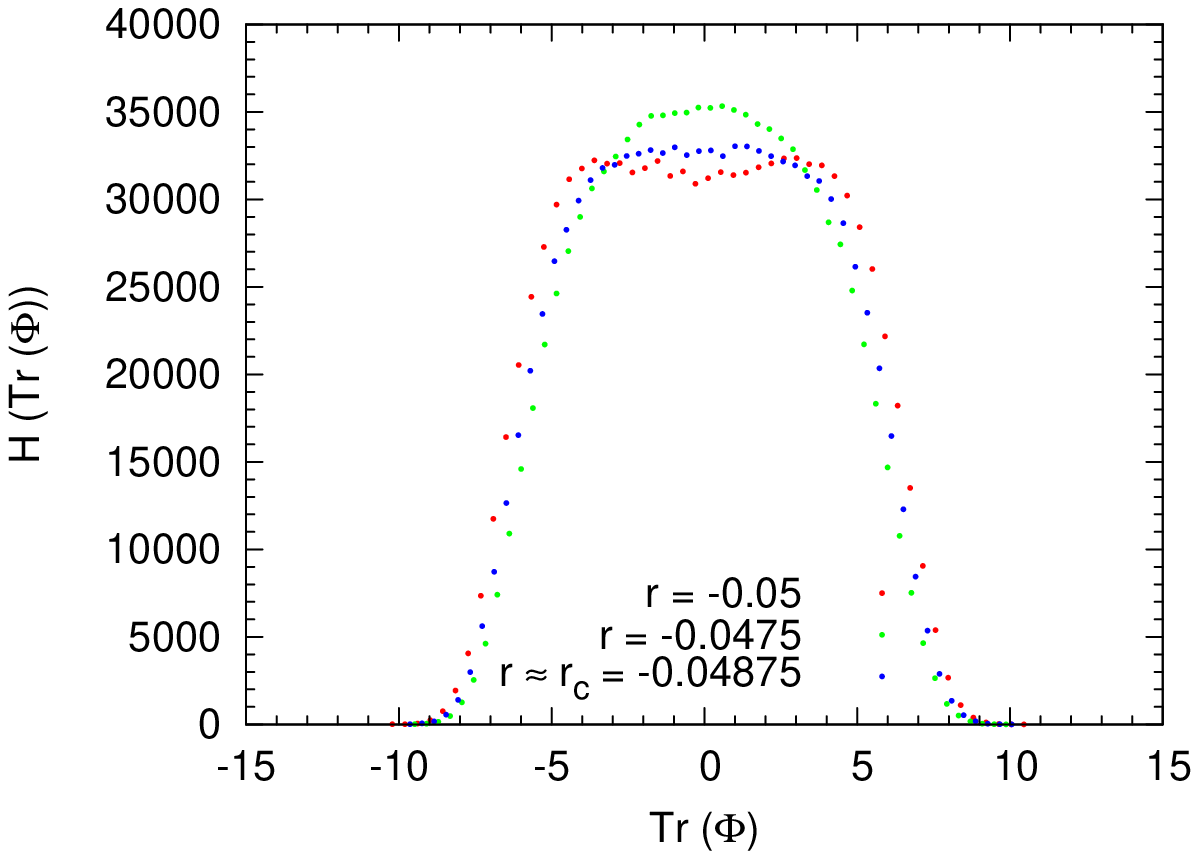}}
\subfigure[\; Distribution of $\Phi_{11}$]
{\label{fig3b}\includegraphics[scale=0.595]{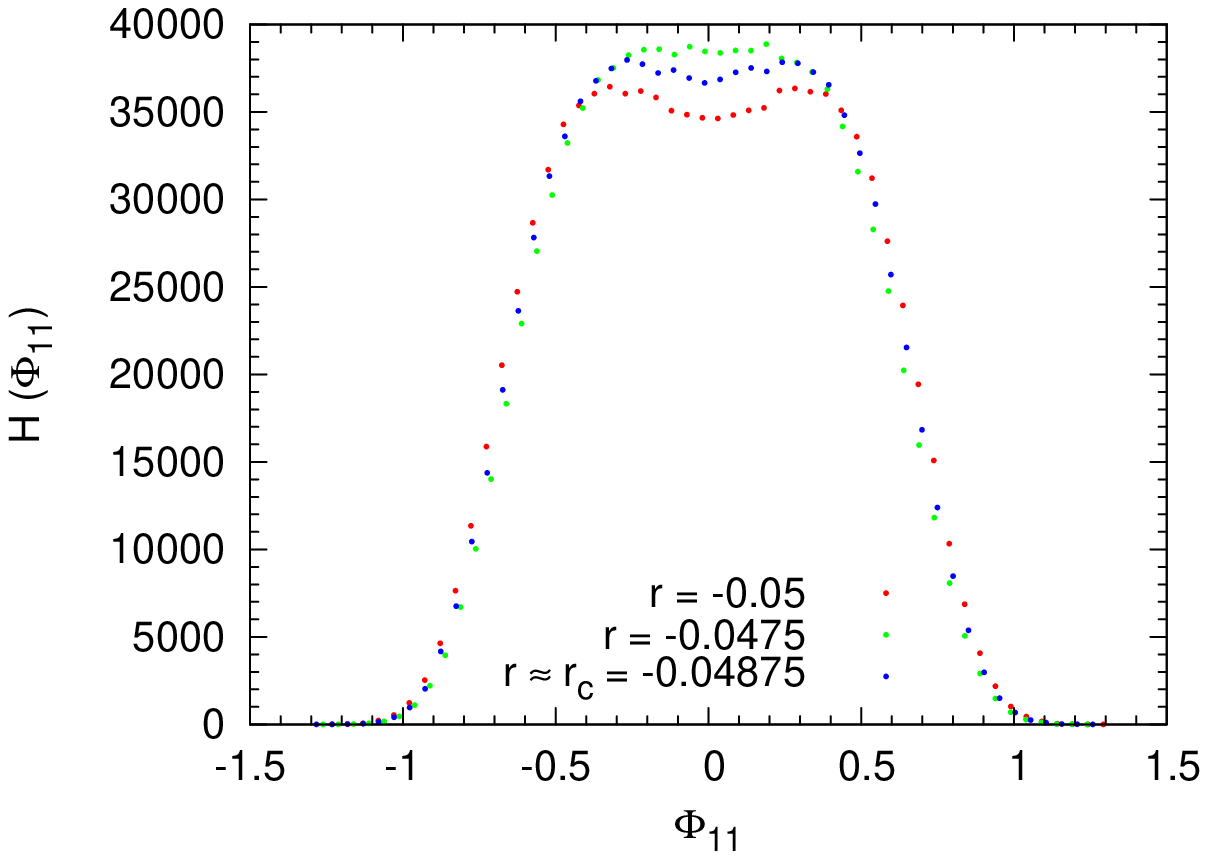}}
\caption{}
\label{fig3}
\end{center}
\end{figure}
\vskip0.25cm
\noindent{\bf Meta-stable states}\\

We also studied cases with very large values of $\lambda R^2$. When $r$ is large 
negative we find different average values for $Tr(\Phi)$ for different initial 
choices of $\Phi$. These different values correspond to local and global minima 
of the effective action. The barrier between these states inhibits the transition
amongst them. The number of these states which we observed 
grow with $N$. This can be more or less seen from the analysis of the action 
itself as fluctuations are not much important for small $r$. The state with
highest $Tr(\Phi)$ found to satisfy $\langle \Phi \rangle \propto \ide$ and 
has the lowest value for the action, hence is the ground state.  
So we conclude that the state with largest $Tr(\Phi)$ is the global minimum
of the system. Other states, which are basically the non-uniform phases,
are meta-stable. We think that the meta-stability
increases with decrease in the average of $Tr(\Phi)$. In Fig.~8(a) we show a 
brief Monte Carlo history of $\Phi$, after thermalization, for different 
initial $\Phi$'s. For higher $r$ the bands persist but move slowly towards 
zero. After certain value of critical $r$ most of these states are observed 
only for sometime in the Monte Carlo history and then the values of 
$Tr(\Phi)$ jumped to zero as seen in Fig.~8(b). 

\begin{figure}[hbt!]
\begin{center}
\subfigure[\; Monte Carlo history of $Tr(\Phi)$ at
$N=16$, $R=15.0$, $\lambda=0.7$ and $r=-3.0$.]
{\label{fig3a}\includegraphics[scale=0.595]{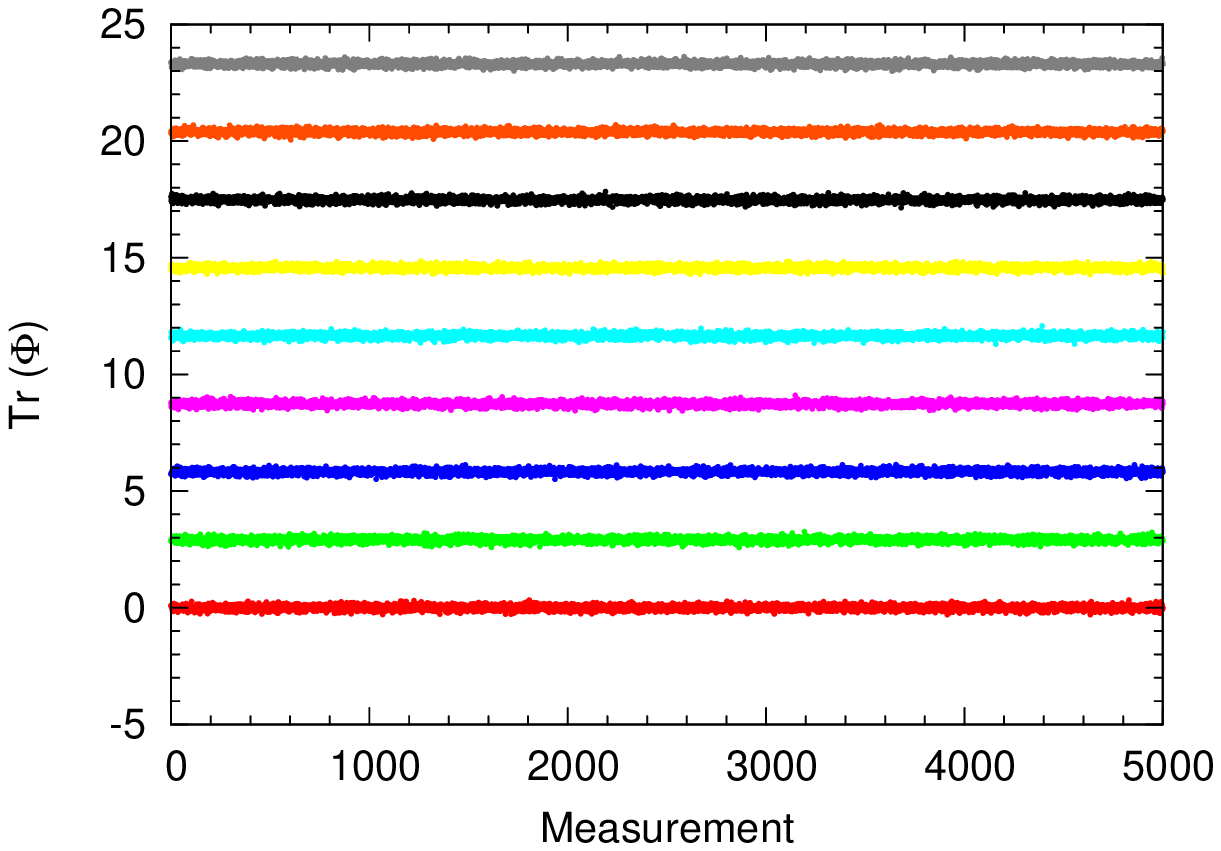}}
\subfigure[\; Monte Carlo history of $Tr(\Phi)$ at
$N=16$, $R=15.0$, $\lambda=0.7$ and $r=-0.885$.]
{\label{fig3b}\includegraphics[scale=0.595]{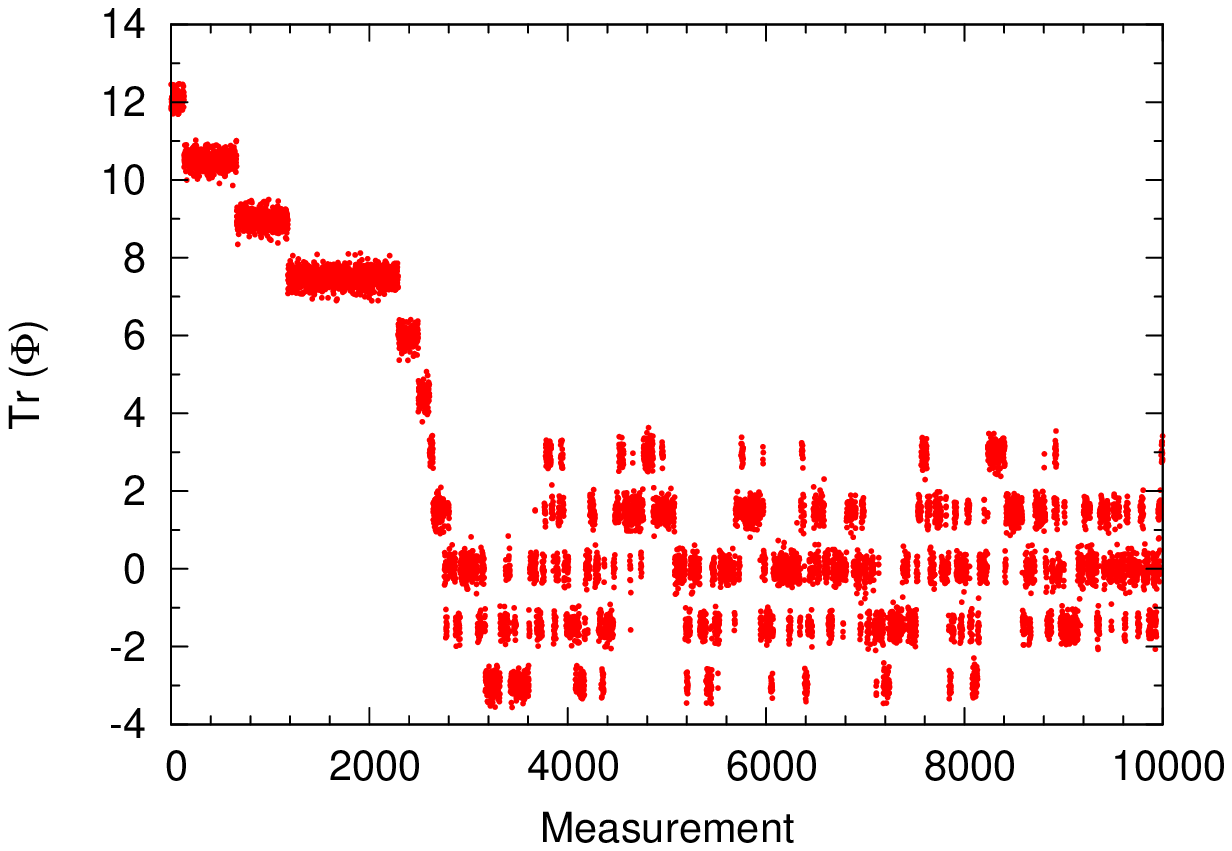}}
\caption{Monte Carlo history for $r < r_1$ and for $r \sim r_1$}
\label{fig3}
\end{center}
\end{figure}

\section{Conclusions}
We have developed a ``pseudo-heatbath'' algorithm to study the finite
temperature phase transitions of $\Phi^4$ theory on a fuzzy sphere.
The results from Monte Carlo simulations clearly show finite temperature
transitions. For some range of $\lambda R^2$, in particular, for
large values one clearly sees stable non-uniform phases for some intermediate 
temperature, intermediate values of $r$. The various phases
are characterised by different properties of $\Phi$. In
the ordered phase this behaves like a identity matrix. All non-uniform
phases have zero $Tr(\Phi)$. Their existence is confirmed by
the peak in the fluctuation of $Tr(\Phi^2)$.
$Tr(\Phi)$ serves as an order parameter for the 
order $\leftrightarrow$ non-uniform transition 
while $\Phi_{11}$ and $\Phi_{NN}$
describe the non-uniform $\leftrightarrow$ disorder transition.

The order-non-uniform transition is found to be first order.
This transition was found to be strong first
order for larger values of $\lambda R^2$.
We conjecture that the first order nature of the transition has to do
with the presence of meta-stable states discussed above. In fact the
state with $Tr(\Phi)=0$ is meta-stable for small temperatures when
$\Phi \propto \ide$ is the absolute ground state. Fluctuations can
only stabilise if it is a stable configuration at higher temperature,
so there is always a barrier with the ordered phase, leading to
first order transition. For smaller values of $\lambda R^2$ the
barrier between the stable and meta-stable phases is not high so
thermal fluctuations make $\Phi$ hop between the different
states. In this case we rather study the distribution of $Tr(\Phi)$
to infer the transition value of $r$ for the order $\leftrightarrow$ non-uniform
transition. From the distribution we find that ground state is 
discontinuously changing. Moreover the distribution of
$Tr(\Phi)$ is very non-gaussian, rather flat near zero, suggesting
that there are degenerate states characterised by zero and
non-zero values $Tr(\Phi)$. So from our results the
transition of ordered phase to other phases is always first order
for the parameter space we have explored.
All other transition in model appeared to
be continuous transitions.

The results from previous studies have shown that by doing simple
scaling phase boundaries for different $N$ coincide \cite{denjoe}.
This expected scaling with $N$ does not occur up to the value of
$N$ we have studied, though it is the largest so far. We studied
the behavior of the triple point for large $N$ and it scales 
approximately linearly with $N$.
However our results seem to agree with 
the previous studies in that the non-uniform phase survives the
continuum limit. 
In our analysis we considered primarily $Tr(\Phi)$, $Tr(\Phi^2)$, 
$\Phi_{11}$ etc.. However analysis of the full matrix may result in better 
understanding of the phase structure, such as variants
of the non-uniform ordered phases.

\section{Acknowledgement}
We thank D. O'Connor, M. Panero, X. Martin, W. Bietenholz for critical 
comments on the draft of this paper. All the calculations have been carried 
out using the computer facilities at IMSc, Chennai.

\end{document}